\documentclass[20pt]{article}
\usepackage{amsmath}
\usepackage{amssymb}
\usepackage{latexsym}
\usepackage{amsmath}
\usepackage{lineno}
\usepackage{graphicx}
\usepackage{float}
\usepackage{subfigure}
\usepackage{geometry}
\usepackage{graphics}
\usepackage{subfigure}
\usepackage{graphicx}
\usepackage{multirow}
\usepackage{booktabs}
\usepackage{hyperref}
\usepackage{cite}
\makeatletter

\newcommand{\be}{\begin{equation}}
\newcommand{\ee}{\end{equation}}

\begin{document}
\begin{center}
\large {\bf Thermodynamics and Phase transition of Schwarzschild black hole in Gravity's Rainbow}
\end{center}

\begin{center}
Zhong-Wen Feng $^1$
 $\footnote{E-mail:  zwfengphy@163.com}$
Shu-Zheng Yang $^{2}$
 $\footnote{E-mail:  szyangcwnu@126.com}$
Hui-Ling Li $ ^{1,3}$
 $\footnote{E-mail:  LHL51759@126.com}$
 Xiao-Tao Zu ${^1}$
 $\footnote{E-mail:  xtzu@uestc.edu.cn}$
\end{center}

\begin{center}
\textit{1. School of Physical Electronics, University of Electronic Science and Technology of China, Chengdu, 610054, China\\
2. Department of Astronomy, China West Normal University, Nanchong, 637009, China\\
3. College of Physics Science and Technology, Shenyang Normal University, Shenyang, 110034, China}
\end{center}

\noindent
{\bf Abstract:} The Planck length and Planck energy should be taken as invariant scales are in agreement with various theories of quantum gravity. In this scenario, the original general relativity can be changed to the so-called gravity's rainbow which produces significant modifications to the black holes' evolution. In this paper, using two kinds of rainbow functions, we investigate the thermodynamics and the phase transition of Schwarzschild black hole in the context of gravity's rainbow theory. Firstly, with the help of the surface gravity and Heisenberg uncertainty principle, we calculate the modified Hawking temperature. Then, based on this modification, we derive the local temperature, free energy and other thermodynamic quantities in an isothermal cavity. Finally, the critical behavior, the thermodynamic stability and phase transition of the rainbow Schwarzschild black hole are analyzed. It turns out that our results are different from those of Hawking-Page phase transition. Meanwhile, we find that there are many similarities and differences between rainbow functions I case and rainbow functions II.

\section{Introduction}
\label{Int}
The Lorentz symmetry is known as one of fundamental symmetries in nature, which produces the standard energy-momentum dispersion relation, $E^2  - p^2  = m^2$\cite{ch1,ch2}. But a lot of works showed that the standard energy-momentum dispersion relation would not be held when it nears the Planck length (or the Planck energy) \cite{ch3,ch4,ch5,ch6,ch7}. In fact, one of the intriguing predictions among various quantum gravity theories, such as loop quantum gravity, loop quantum gravity and non-commutative geometry, is the existence of a minimum measurable length that can be identified with the Planck scale \cite{ch8,ch9,ch10}. This idea is supported by many Gedanken experiments \cite{ch11}. Furthermore, the Planck length can be considered as the division between the quantum and the classical description of spacetime \cite{ch12}. Therefore, the Planck length should be taken as an invariant scale, that is, a universal constant for all inertial observers. However, in the special relativity, those features are conflict with the Lorentz symmetry since the Planck scale is not invariant under the linear Lorentz transformations. In order to solve this paradoxical situation, the standard energy-momentum dispersion relation must be changed to the so-called modified dispersion relation (MDR). The MRD plays an important role in physics \cite{ch13,ch14,ch15,ch16,ch89a+,ch89b+,ch17,ch18,ch19,ch20}. In Refs.~\cite{ch13,ch14,ch15,ch16}, authors elucidated that the MRD may be responsible for the threshold anomalies of in ultra-high-energy cosmic rays and TeV photons. Besides, the MDR provides stringent constraints on deformations of special relativity and Lorentz violations \cite{ch89a+,ch89b+,ch17,ch18,ch19,ch20}. Very recently, Arzano and Calcagni showed that the MRD may affect the propagation of the observed gravitational-wave signal of the event GW 150914 \cite{ch20+,ch21+}.

Combining the MDR with the special relativity, Amelino-Camelia proposed the double special relativity (DSR) \cite{ch21}. The DSR is an extension of special relativity, which contains two fundamental constants: one is the velocity of light  $c$, the other is the Planck energy $E_p$, they indicate that the energy and the velocity of a particle cannot exceed the velocity of light and the Planck energy. Meanwhile, DSR is also a framework for encoding properties of flat quantum spacetime \cite{ch22}. However, it is observed that DSR is typically formulated in momentum space, it leads to the definition of a dual space suffers a nonlinearity of the Lorentz transformation, that is, the dual space is non-trivial. In order to overcome this problem, the DSR was generalized to the curved spacetimes by Magueijo and Smolin. This doubly general theory of relativity is called as gravity's rainbow (or rainbow gravity) \cite{ch23}. The name gravity's rainbow comes from the fact that this theory assumes the spacetime background depends on the energy of a test particle. Therefore, one should use a family of metrics (namely, a rainbow of metrics) parameterized by the ratio ${E \mathord{\left/{\vphantom {E {E_p }}} \right.\kern-\nulldelimiterspace} {E_p }}$  to describe the background of this spacetime instead a single metric. In the gravity's rainbow, using a modified equivalence principle \cite{ch23}, the modified metrics can be expressed as $\hat g = \eta ^{\mu \nu } e_\mu  \left( {{E \mathord{\left/ {\vphantom {E {E_p }}} \right. \kern-\nulldelimiterspace} {E_p }}} \right) \otimes e_\nu  \left( {{E \mathord{\left/ {\vphantom {E {E_p }}} \right. \kern-\nulldelimiterspace} {E_p }}} \right)$ with the energy dependence of the frame fields $e_0 \left( {{E \mathord{\left/ {\vphantom {E {E_p }}} \right. \kern-\nulldelimiterspace} {E_p }}} \right) = {{\hat e_0 } \mathord{\left/ {\vphantom {{\hat e_0 } {f\left( {{E \mathord{\left/ {\vphantom {E {E_p }}} \right. \kern-\nulldelimiterspace} {E_p }}} \right)}}} \right. \kern-\nulldelimiterspace} {f\left( {{E \mathord{\left/ {\vphantom {E {E_p }}} \right. \kern-\nulldelimiterspace} {E_p }}} \right)}}$ and $e_i \left( {{E \mathord{\left/ {\vphantom {E {E_p }}} \right. \kern-\nulldelimiterspace} {E_p }}} \right) = {{\hat e_i } \mathord{\left/ {\vphantom {{\hat e_i } {f\left( {{E \mathord{\left/ {\vphantom {E {E_p }}} \right. \kern-\nulldelimiterspace} {E_p }}} \right)}}} \right. \kern-\nulldelimiterspace} {f\left( {{E \mathord{\left/ {\vphantom {E {E_p }}} \right. \kern-\nulldelimiterspace} {E_p }}} \right)}}$. Due to the modified metrics, the connection and curvature become energy dependent, and the Einstein's equations are modified as  $G_{\mu \nu } \left( {{E \mathord{\left/ {\vphantom {E {E_p }}} \right. \kern-\nulldelimiterspace} {E_p }}} \right) = 8\pi G\left( {{E \mathord{\left/ {\vphantom {E {E_p }}} \right. \kern-\nulldelimiterspace} {E_p }}} \right)T_{\mu \nu } \left( {{E \mathord{\left/ {\vphantom {E {E_p }}} \right. \kern-\nulldelimiterspace} {E_p }}} \right) + g_{\mu \nu } \lambda \left( {{E \mathord{\left/ {\vphantom {E {E_p }}} \right. \kern-\nulldelimiterspace} {E_p }}} \right)$, where the  $G\left( {{E \mathord{\left/ {\vphantom {E {E_p }}} \right. \kern-\nulldelimiterspace} {E_p }}} \right)$ is the effective gravitational coupling constant at the ultraviolet (UV) regime.

The gravity's rainbow is very important since it can be applied to different physical systems and modifies many classical theories. For example, by incorporating the gravity's rainbow with the FRW cosmologies, the possibility of resolving big bang singularity has been investigated in Ref.~\cite{ch24}. In Ref.~\cite{ch25}, Sefiedgar studied the entropic force in the context of gravity's rainbow. Then, based on this entropic force, the modified the Newtonian dynamics and Einstein's field equations were obtained. In Ref.~\cite{ch26}, with the help of gravity's rainbow, the authors investigated the modified Starobinsky model and the inflationary solution to the motion equations, the spectral index of curvature perturbation and the tensor-to-scalar ratio were also calculated. Besides, the deflection of light, photon time delay, gravitational red-shift, and the weak equivalence principle (WEP) have also been studied in the framework of gravity's rainbow \cite{ch27}. It is worth mentioning that the gravity's rainbow has a great influence on the thermodynamics of black holes \cite{ch28,ch29,ch30,ch31,ch32,ch33,ch34,ch35,ch36,ch36+,ch37+,ch37,ch38,ch39,ch40,ch41,ch42}. In 2004, Ali investigated the thermodynamics of rainbow Schwarzschild (SC) black hole by using the gravity's rainbow and the Heisenberg uncertainty principle (HUP) \cite{ch31}. The results showed that the mass-temperature relation of rainbow SC black hole is different from that of the original case. As the mass of the black hole decreases, the modified Hawking temperature reaches a maximum value, and then, it goes to zero when the mass of rainbow SC black hole approaches the Planck scale. Subsequently, similar calculations have been applied to other black holes \cite{ch32,ch33,ch34,ch35,ch36,ch36+,ch37+,ch37,ch38,ch39,ch40,ch41,ch42}. Those results implies that the gravity's rainbow can prevent the black holes from total evaporation and leads to the remnants of black holes in exactly the same way as done by the generalized uncertainty principle (GUP) \cite{ch43}. Hence,  the gravity's rainbow may solve the information loss and naked singularity problems of black holes.

On the other hand, it is well known that black holes not only have the thermodynamic quantities, but also have rich phase structures and critical phenomena. Over the past decades, the phase transitions and critical phenomena of black holes have been widely explored \cite{ch43+}. In 1977, Davies found that the Kerr-Newman black hole exists the phenomena of phase transition \cite{ch44}. Then, Pav\'{o}n discovered a non-equilibrium second order phase transition in the charged Reissner-Nordstr\"{o}m (RN) black hole spacetime \cite{ch45}. The existence of a certain phase transition in the asymptotically anti de Sitter (AdS) spacetimes was proved by Hawking and Page. In Ref.~\cite{ch46}, they demonstrated that the AdS SC black hole undergoes a phase transition (namely, Hawking-Page phase transition) to a thermal AdS space if the temperature reaches a certain value. This seminal work has attracted wide attention because it can explain the confinement/deconfinement phase transition of gauge field in the AdS/CFT correspondence \cite{ch47}. Since then, a lot of works have been devoted to investigate the Hawking-Page phase transition and the critical phenomena for other more complicated AdS spacetimes \cite{ch47,ch48,ch49,ch50,ch51,ch52,ch53,ch54,ch55,ch56,ch57}. Among those studies, people found that the phase transitions behavior of charged black holes are similar to that of Van der Waals liquid-gas system \cite{ch50,ch51,ch52,ch53,ch54,ch55,ch56,ch57}. In 2012, Kubiz\v{n}\'{a}k and Mann proposed a remarkable new perspective on relation between the phase transition of charged AdS black hole and the van der Waals liquid-gas phase transition \cite{ch50}. By considering the cosmological constant $\lambda$ and its conjugate variable as thermodynamic pressure $P$  and specific volume $V$, they analyzed the thermodynamic behavior of RN-AdS black hole in the extended phase space. The results showed that the RN-AdS black hole system has a first-order small-large black hole phase transition since the free energy $G$ demonstrates a ``swallow tail''-type behavior. Besides, Kubiz\v{n}\'{a}k and Mann also studied the ``$P-V$''  criticality and the critical exponents and proved that they coincide with those of the Van der Waals system.

Based on what has been discussed above, it is interesting to raise a question whether it is possible to investigate the thermodynamic criticality and the phase transition of the rainbow black holes? In Refs.~\cite{ch36,ch37,ch38,ch39,ch40,ch41,ch42,ch58,ch59,ch59+}, the authors proved that it is indeed possible. Very recently, Gim and Kim have investigated the thermodynamic phase transition of a black hole in the context of gravity's rainbow. Their results showed that the rainbow SC black hole in an isothermal cavity an additional Hawking-Page-type critical temperature near the event horizon apart from the original case, which is of relevance to the existence of a locally small black hole. Meanwhile, by analyzing the free energy, they concluded that the small black hole will eventually tunnel into the stable large black hole since it just locally stable \cite{ch59}. Motivated by Refs.~\cite{ch26,ch59,ch59+}, we would like to investigate the thermodynamic criticality and phase transition of rainbow SC black hole. By utilizing the Heisenberg uncertainty principle, the value of the energy of emitted particle $E$ can be taken to be the mass of black hole $M$, that is  $E \ge {1 \mathord{\left/ {\vphantom {1 {r_H }}} \right. \kern-\nulldelimiterspace} {r_H }} = {1 \mathord{\left/ {\vphantom {1 {2GM}}} \right. \kern-\nulldelimiterspace} {2GM}}$. According to this relation and two proposals for rainbow functions, we derive the modified thermodynamic quantities and free energy of the rainbow SC black hole. Based on those modifications, the thermodynamic criticality and phase transition are analyzed. It turns out that our results are different from the those of Hawking-Page phase transition. Meanwhile, it is shown that there are many similarities and differences between rainbow functions I case and rainbow functions II case. Moreover, our results are also different from that in Ref.~\cite{ch59}. However, when the ratio ${E \mathord{\left/{\vphantom {E {E_p }}} \right.\kern-\nulldelimiterspace} {E_p }}$ approaches zero, our results reduce to the standard forms.

The organization of this paper is as follows. The next section is devoted to introducing two kinds of rainbow functions. In section \ref{H-J}, according to the line element of rainbow SC black hole, we will use the Hamilton-Jacobi method to compute the modified surface gravity and the effective temperature. In section \ref{TP I}, using the rainbow functions defined in Ref.~\cite{ch13}, we study the thermodynamics of SC black hole in the context of gravity's rainbow and discuss its thermodynamic criticality and phase transition. In section \ref{TP II}, we use the other kinds of rainbow functions to study the thermodynamics, critical phenomena and phase transition of rainbow SC black hole. The paper ends with conclusions in Section \ref{Dis}.

\section{A brief about the rainbow functions}
\label{RF}
To begin with, we review briefly the MRD and the rainbow functions. The general form of MRD can be expressed as follows
\begin{equation}
\label{eq1}
E^2 f^2 \left( {{E \mathord{\left/ {\vphantom {E {E_p }}} \right. \kern-\nulldelimiterspace} {E_p }}} \right) - p^2 g^2 \left( {{E \mathord{\left/
 {\vphantom {E {E_p }}} \right. \kern-\nulldelimiterspace} {E_p }}} \right) = m^2 ,
\end{equation}
where  $E_p$ is the Planck energy, the corrections terms  $f\left( {{E \mathord{\left/{\vphantom {E {E_p }}} \right. \kern-\nulldelimiterspace} {E_p }}} \right)
$ and $g\left( {{E \mathord{\left/ {\vphantom {E {E_p }}} \right. \kern-\nulldelimiterspace} {E_p }}} \right)$  are known as rainbow functions which are responsible for the modification of the energy-momentum relation at UV regime. However, in the limit  ${E \mathord{\left/ {\vphantom {E {E_p }}} \right. \kern-\nulldelimiterspace} {E_p }} \to 0$, the rainbow functions satisfy the relations  $\mathop {\lim }\limits_{{E \mathord{\left/ {\vphantom {E {E_p }}} \right. \kern-\nulldelimiterspace} {E_p }} \to 0} f\left( {{E \mathord{\left/ {\vphantom {E {E_p }}} \right. \kern-\nulldelimiterspace} {E_p }}} \right) = 1
$ and  $\mathop {\lim }\limits_{{E \mathord{\left/ {\vphantom {E {E_p }}} \right. \kern-\nulldelimiterspace} {E_p }} \to 0} g\left( {{E \mathord{\left/
 {\vphantom {E {E_p }}} \right. \kern-\nulldelimiterspace} {E_p }}} \right) = 1$, the relations indicate that the MDR will reduce to the standard energy-momentum dispersion relation at low energy scale.

It should be mentioned that the expression of rainbow function is not unique, people can find a series expressions of rainbow functions based on different phenomenological motivations. In Refs.~\cite{ch60,ch61}, motivated by the results of non-critical string theory,  loop quantum gravity and  $\kappa$-Minkowski non-commutative geometry, Amelino-Camelia and Ellis proposed one of the most studied MDR which is $E^2  - p^2  + \eta p^2 \left( {{E \mathord{\left/ {\vphantom {E {E_p }}} \right. \kern-\nulldelimiterspace} {E_p }}} \right)^n  = m^2$. Comparing the MRD with Eq.~(\ref{eq1}), the corresponding rainbow functions can be expressed as
\begin{equation}
\label{eq2}
   {f\left( {{E \mathord{\left/
 {\vphantom {E {E_p }}} \right.
 \kern-\nulldelimiterspace} {E_p }}} \right) = 1,} \quad {g\left( {{E \mathord{\left/
 {\vphantom {E {E_p }}} \right.
 \kern-\nulldelimiterspace} {E_p }}} \right) = \sqrt {1 - \eta \left( {{E \mathord{\left/
 {\vphantom {E {E_p }}} \right.
 \kern-\nulldelimiterspace} {E_p }}} \right)^n } }  \\,
\end{equation}
which we identify it as first  RFs I.  The $\eta$ is a positive free parameter that represents the rainbow parameter, and $n$ represents an integer. Very recently, according to a varying speed of light theory (VSL), the authors in Ref.~\cite{ch13} constructed a kind of MRD, which takes the form as ${{E^2 } \mathord{\left/  {\vphantom {{E^2 } {\left( {1 - {{\gamma E} \mathord{\left/ {\vphantom {{\gamma E} {E_p }}} \right. \kern-\nulldelimiterspace} {E_p }}} \right)^2 }}} \right.  \kern-\nulldelimiterspace} {\left( {1 - {{\gamma E} \mathord{\left/ {\vphantom {{\gamma E} {E_p }}} \right. \kern-\nulldelimiterspace} {E_p }}} \right)^2 }} - p^2  = m^2$. This kind of MRD implies a spacetime has the energy-dependent velocity $c = \left( {1 - {{\gamma E} \mathord{\left/ {\vphantom {{\gamma E} {E_p }}} \right. \kern-\nulldelimiterspace} {E_p }}} \right){}$. Comparing it with Eq.~(\ref{eq1}), the rainbow functions can be fixed as follows
\begin{equation}
\label{eq3}
f\left( {{E \mathord{\left/
 {\vphantom {E {E_p }}} \right.
 \kern-\nulldelimiterspace} {E_p }}} \right) = \frac{1}{{1 - {{\gamma E} \mathord{\left/
 {\vphantom {{\gamma E} {E_p }}} \right.
 \kern-\nulldelimiterspace} {E_p }}}}, \quad  g\left( {{E \mathord{\left/
 {\vphantom {E {E_p }}} \right.
 \kern-\nulldelimiterspace} {E_p }}} \right) = 1,
\end{equation}
where $\gamma$  is the rainbow parameter. Eq.~(\ref{eq3}) implies that the varying velocity of light in the rainbow gravity becomes smaller when the energy of photons increases. For convenience, we call Eq.~(\ref{eq3}) the RFs II. Notably, the bounds on the values of $\eta$ and $\gamma$ have been analyzed by using many theoretical and experimental considerations \cite{ch27}. The concrete expressions of rainbow functions have a strong influence on the predictions. Hence, in the subsequent discussions, we will calculate the thermodynamics of SC black hole taking into account the effect of the RF I and RF II. Then, the phase transition and critical phenomena in terms of gravity's rainbow will be analyzed.

\section{Rainbow SC black hole and the Hamilton-Jacobi method}
\label{H-J}
In Refs.~\cite{ch23,ch43}, the authors pointed out that one can construct the rainbow spacetime by making replacements $dt \to {{dt} \mathord{\left/ {\vphantom {{dt} {f\left( {{E \mathord{\left/ {\vphantom {E {E_p }}} \right. \kern-\nulldelimiterspace} {E_p }}} \right)}}} \right. \kern-\nulldelimiterspace} {f\left( {{E \mathord{\left/ {\vphantom {E {E_p }}} \right. \kern-\nulldelimiterspace} {E_p }}} \right)}}$  for time coordinates and $dx^i  \to {{dx^i } \mathord{\left/ {\vphantom {{dx^i } {g\left( {{E \mathord{\left/ {\vphantom {E {E_p }}} \right. \kern-\nulldelimiterspace} {E_p }}} \right)}}} \right. \kern-\nulldelimiterspace} {g\left( {{E \mathord{\left/ {\vphantom {E {E_p }}} \right. \kern-\nulldelimiterspace} {E_p }}} \right)}}$  for all spatial coordinates. Therefore, the line element of rainbow Schwarzschild black hole is given by
\begin{equation}
\label{eq4}
ds^2  =  - \frac{{1 - \left( {{{2GM} \mathord{\left/
 {\vphantom {{2GM} r}} \right.
 \kern-\nulldelimiterspace} r}} \right)}}{{f^2 \left( {{E \mathord{\left/
 {\vphantom {E {E_p }}} \right.
 \kern-\nulldelimiterspace} {E_p }}} \right)}}dt^2  + \frac{{dr^2 }}{{\left[ {1 - \left( {{{2GM} \mathord{\left/
 {\vphantom {{2GM} r}} \right.
 \kern-\nulldelimiterspace} r}} \right)} \right]g^2 \left( {{E \mathord{\left/
 {\vphantom {E {E_p }}} \right.
 \kern-\nulldelimiterspace} {E_p }}} \right)}} + \frac{{r^2 }}{{g^2 \left( {{E \mathord{\left/
 {\vphantom {E {E_p }}} \right.
 \kern-\nulldelimiterspace} {E_p }}} \right)}}d\Omega ^2 ,
\end{equation}
where $d\Omega ^2  = d\theta ^2  + \sin ^2 \theta d\phi ^2$  represents the line elements of 2-dimensional hypersurfaces. Obviously, the event horizon of rainbow SC black hole is located at  $r_H  = 2GM$. In the limit ${E \mathord{\left/ {\vphantom {E {E_p }}} \right. \kern-\nulldelimiterspace} {E_p }} \to 0$, the metric of original SC black hole is recovered. In order to investigate the modified Hawking temperature of rainbow SC black hole, we can compute the tunneling behaviors of particles by using the Hamilton-Jacobi method. In Ref.~\cite{ch43,ch66+}, the authors derived the deformed Hamilton-Jacobi equations $g_{\mu \nu } \left( {\partial ^\mu  I} \right)\left( {\partial ^\nu  I} \right) = 0$ for scalars, spin $1/2$ fermions and vector bosons in the line element~(\ref{eq4}), where $I$ is the action of a particle. Therefore, the Hamilton-Jacobi equation for a massless particle in the rainbow metric is given by
\begin{eqnarray}
\label{eq4a+}
 - \frac{{f^2 \left( {{E \mathord{\left/
 {\vphantom {E {E_p }}} \right.
 \kern-\nulldelimiterspace} {E_p }}} \right)}}{{1 - \left( {{{2GM} \mathord{\left/
 {\vphantom {{2GM} r}} \right.
 \kern-\nulldelimiterspace} r}} \right)}}\left( {\partial _t I} \right)^2  + g^2 \left( {{E \mathord{\left/
 {\vphantom {E {E_p }}} \right.
 \kern-\nulldelimiterspace} {E_p }}} \right)\left[ {1 - \left( {{{2GM} \mathord{\left/
 {\vphantom {{2GM} r}} \right.
 \kern-\nulldelimiterspace} r}} \right)} \right]\left( {\partial _r I} \right)^2  + \frac{{g^2 \left( {{E \mathord{\left/
 {\vphantom {E {E_p }}} \right.
 \kern-\nulldelimiterspace} {E_p }}} \right)}}{{r^2 }}\left[ {\left( {\partial _\theta  I} \right)^2  + \frac{{\left( {\partial _\phi  I} \right)^2 }}{{\sin ^2 \theta }}} \right] = 0.
\end{eqnarray}
Taking into account the time-like killing vector $\left( {{\partial  \mathord{\left/  {\vphantom {\partial  {\partial t}}} \right. \kern-\nulldelimiterspace} {\partial t}}} \right)^a$ and the the space-like killing vector  $\left( {{\partial  \mathord{\left/  {\vphantom {\partial  {\partial \phi}}} \right. \kern-\nulldelimiterspace} {\partial \phi}}} \right)^a$, one can employ the ansatz $I =  - \omega t + W\left( r \right) + \Theta \left( \theta,  \right) +j \phi $ with the energy of particle $\omega$. Inserting this ansatz into Eq.~(\ref{eq4a+}), yielding
\begin{eqnarray}
\label{eq4b+}
 - f^2 \left( {{E \mathord{\left/
 {\vphantom {E {E_p }}} \right.
 \kern-\nulldelimiterspace} {E_p }}} \right)\left( {1 - {{2GM} \mathord{\left/
 {\vphantom {{2GM} r}} \right.
 \kern-\nulldelimiterspace} r}} \right)^{ - 1} \omega ^2  + g^2 \left( {{E \mathord{\left/
 {\vphantom {E {E_p }}} \right.
 \kern-\nulldelimiterspace} {E_p }}} \right)\left( {1 - {{2GM} \mathord{\left/
 {\vphantom {{2GM} r}} \right.
 \kern-\nulldelimiterspace} r}} \right)\left( {\partial _r W} \right)^2  =  - {\lambda  \mathord{\left/
 {\vphantom {\lambda  {r^2 }}} \right.
 \kern-\nulldelimiterspace} {r^2 }},
\end{eqnarray}
\begin{eqnarray}
\label{eq4c+}
g^2 \left( {{E \mathord{\left/
 {\vphantom {E {E_p }}} \right.
 \kern-\nulldelimiterspace} {E_p }}} \right)\left[ {\left( {\partial _\theta  \Theta } \right)^2  + j^2 \sin ^{ - 2} \theta  } \right] = \lambda,
\end{eqnarray}
where $\lambda$ is a constant. When considering the $s$-wave, it is found that the angle part of the Hamilton-Jacobi equation dose not contribute to the tunneling behaviors. Therefore, solving  Eq.~(\ref{eq4b+}), one has
\begin{eqnarray}
\label{eq4d+}
W\left( r \right) = \pm \int {  \left( {1 - {{2GM} \mathord{\left/ {\vphantom {{2GM} r}} \right. \kern-\nulldelimiterspace} r}} \right)^{ - 1} \sqrt {{{f^2 \left( {{E \mathord{\left/ {\vphantom {E {E_p }}} \right. \kern-\nulldelimiterspace} {E_p }}} \right)\omega ^2 } \mathord{\left/ {\vphantom {{f^2 \left( {{E \mathord{\left/ {\vphantom {E {E_p }}} \right. \kern-\nulldelimiterspace} {E_p }}} \right)\omega ^2 } {g^2 \left( {{E \mathord{\left/ {\vphantom {E {E_p }}} \right. \kern-\nulldelimiterspace} {E_p }}} \right)}}} \right. \kern-\nulldelimiterspace} {g^2 \left( {{E \mathord{\left/ {\vphantom {E {E_p }}} \right.
 \kern-\nulldelimiterspace} {E_p }}} \right)}} - {\lambda  \mathord{\left/ {\vphantom {\lambda  {\left( {1 - {{2GM} \mathord{\left/ {\vphantom {{2GM} r}} \right. \kern-\nulldelimiterspace} r}} \right)r^2 }}} \right. \kern-\nulldelimiterspace} {\left( {1 - {{2GM} \mathord{\left/ {\vphantom {{2GM} r}} \right. \kern-\nulldelimiterspace} r}} \right)r^2 }}} } dr,
\end{eqnarray}
where $+(-)$ are the outgoing (incoming) solutions of the emitted particles. Using the residue theory for the semi circle, the imaginary part of the action for the tunneling process is given by 
\begin{eqnarray}
\label{eq4e+}
{\mathop{\rm Im}\nolimits} W_ \pm  \left( r \right) =  \pm \frac{\omega \pi }{{2\kappa }} = \frac{\omega \pi }{{2\kappa _0 }}\frac{{f\left( {{E \mathord{\left/
 {\vphantom {E {E_p }}} \right.
 \kern-\nulldelimiterspace} {E_p }}} \right)}}{{g\left( {{E \mathord{\left/
 {\vphantom {E {E_p }}} \right.
 \kern-\nulldelimiterspace} {E_p }}} \right)}}.
\end{eqnarray}
From above expression, it is evident that the original surface gravity ${\kappa _0 }= 2 G M$ gets modified in gravity's rainbow. As shown in \cite{ch62,ch63,ch64+}, the particle tunneling rate of rainbow SC black hole is $\Gamma  = {{\Gamma _{{\rm{emit}}} } \mathord{\left/ {\vphantom {{\Gamma _{{\rm{emit}}} } {\Gamma _{{\rm{absorb}}} }}} \right. \kern-\nulldelimiterspace} {\Gamma _{{\rm{absorb}}} }} = {{\exp \left( { - 2{\mathop{\rm Im}\nolimits} W_ +  } \right)} \mathord{\left/ {\vphantom {{\exp \left( { - 2{\mathop{\rm Im}\nolimits} W_ +  } \right)} {\exp \left( { - 2{\mathop{\rm Im}\nolimits} W_ -  } \right)}}} \right. \kern-\nulldelimiterspace} {\exp \left( { - 2{\mathop{\rm Im}\nolimits} W_ -  } \right)}} = \exp \left[ {{{2\omega \pi f\left( {{E \mathord{\left/ {\vphantom {E {E_p }}} \right. \kern-\nulldelimiterspace} {E_p }}} \right)} \mathord{\left/ {\vphantom {{2\omega \pi f\left( {{E \mathord{\left/ {\vphantom {E {E_p }}} \right. \kern-\nulldelimiterspace} {E_p }}} \right)} {\kappa _0 g\left( {{E \mathord{\left/ {\vphantom {E {E_p }}} \right. \kern-\nulldelimiterspace} {E_p }}} \right)}}} \right. \kern-\nulldelimiterspace} {\kappa _0 g\left( {{E \mathord{\left/ {\vphantom {E {E_p }}} \right. \kern-\nulldelimiterspace} {E_p }}} \right)}}} \right]$. With the help of a Boltzmann factor, the effective temperature can be expressed as follows
\begin{equation}
\label{eq5}
T^{RFs}_{eff} = \frac{\kappa }{{2\pi }} =\frac{{\kappa _0 }}{{2\pi }}\frac{{g\left( {{E \mathord{\left/ {\vphantom {E {E_p }}} \right. \kern-\nulldelimiterspace} {E_p }}} \right)}}{{f\left( {{E \mathord{\left/ {\vphantom {E {E_p }}} \right. \kern-\nulldelimiterspace} {E_p }}} \right)}} = T_0 \frac{{g\left( {{E \mathord{\left/ {\vphantom {E {E_p }}} \right. \kern-\nulldelimiterspace} {E_p }}} \right)}}{{f\left( {{E \mathord{\left/ {\vphantom {E {E_p }}} \right.
 \kern-\nulldelimiterspace} {E_p }}} \right)}} = \frac{1}{{8\pi GM}}\frac{{g\left( {{E \mathord{\left/ {\vphantom {E {E_p }}} \right. \kern-\nulldelimiterspace} {E_p }}} \right)}}{{f\left( {{E \mathord{\left/ {\vphantom {E {E_p }}} \right. \kern-\nulldelimiterspace} {E_p }}} \right)}}.
\end{equation}
where $T_0 = {1 \mathord{\left/ {\vphantom {1 {8\pi GM}}} \right. \kern-\nulldelimiterspace} {8\pi GM}}$ is the original temperature. Obviously, Eq.~(\ref{eq5}) shows there is a correction to the effective temperature and the correction value is dependent not only on the mass of black hole but also on the expression of rainbow functions. 

\section{Thermodynamics and phase transition of rainbow SC black hole with rainbow functions I}
\label{TP I}
In this section, we will investigate the thermodynamics and phase transition of rainbow SC black hole with rainbow functions I. Substituting the identification of rainbow functions~(\ref{eq2}) into Eq.~(\ref{eq5}), the modified Hawking temperature can be expressed as \cite{ch31}
\begin{equation}
\label{eq6}
T_H^{RFsI}  = \sqrt {1 - \eta \left( {\frac{E}{{E_p }}} \right)^n }  T_H ,
\end{equation}
where the original Hawking temperature of SC black hole is $T_H  = {1 \mathord{\left/ {\vphantom {1 {8\pi GM}}} \right.\kern-\nulldelimiterspace} {8\pi GM}}
$. According to Refs.~\cite{ch64,ch65}, the Heisenberg uncertainty principle (HUP)  $\Delta x\Delta p \ge 1$  still holds in the gravity's rainbow. Therefore, when considering the photons that in the vicinity of the black hole surface, the HUP can be translated into a lower bound on the energy, that is, $E \ge {1 \mathord{\left/ {\vphantom {1 {\Delta x}}} \right. \kern-\nulldelimiterspace} {\Delta x}}$, where $E$ is the energy of a particle emitted in the Hawking radiation. With the help of lower bound on the energy and the uncertainty position $\Delta x$, one has the following relation
\begin{equation}
\label{eq7}
E \ge {1 \mathord{\left/
 {\vphantom {1 {\Delta x}}} \right.
 \kern-\nulldelimiterspace} {\Delta x}} \approx {1 \mathord{\left/
 {\vphantom {1 {r_H }}} \right.
 \kern-\nulldelimiterspace} {r_H }} = {1 \mathord{\left/
 {\vphantom {1 {2GM}}} \right.
 \kern-\nulldelimiterspace} {2GM}}.
\end{equation}
It should be noted that this bound on the energy plays a key role in the modification of thermodynamics of rainbow black holes. Substituting Eq.~(\ref{eq7}) into the Eq.~(\ref{eq6}), the rainbow Hawking temperature can be rewritten as
\begin{equation}
\label{eq8}
T_H ^{RFsI} = \frac{1}{{4\pi \left( {2GM} \right)^{\frac{{n + 2}}{n}} }}\sqrt {\left( {2GM} \right)^n  - \eta G^{\frac{n}{2}} } ,
\end{equation}
where we use $E_p  = {1 \mathord{\left/ {\vphantom {1 {\sqrt G }}} \right. \kern-\nulldelimiterspace} {\sqrt G }}$ in natural unit. When  $\eta=0$, the rainbow Hawking temperature goes to the original case. Obviously, the rainbow Hawking temperature is very sensitive to the concrete expression of rainbow functions. Next, we will investigate the entropy of rainbow SC black hole. Using the first law of black hole thermodynamics and setting  $n=2$ as an example, the rainbow entropy associated with the rainbow Hawking temperature Eq.~(\ref{eq8}) is given by
\begin{equation}
\label{eq9}
S^{RFsI} = \int {T_H^{ - 1} dM = \pi \left[ {2M\chi  + \eta {\rm{ln}}\left( {2GM + \chi } \right)} \right]} ,
\end{equation}
where $\chi  = \sqrt {G\left( {4GM^2  - \eta } \right)}$. It is clear that the area law of the entropy $S = {A \mathord{\left/ {\vphantom {A 4}} \right.
 \kern-\nulldelimiterspace} 4} = 4\pi M^2$ can be recovered when $\eta=0$, where $A$ is the area of black hole

\begin{figure}
\centering
\subfigure[]{
\begin{minipage}[b]{0.42\textwidth}
\includegraphics[width=1.25\textwidth]{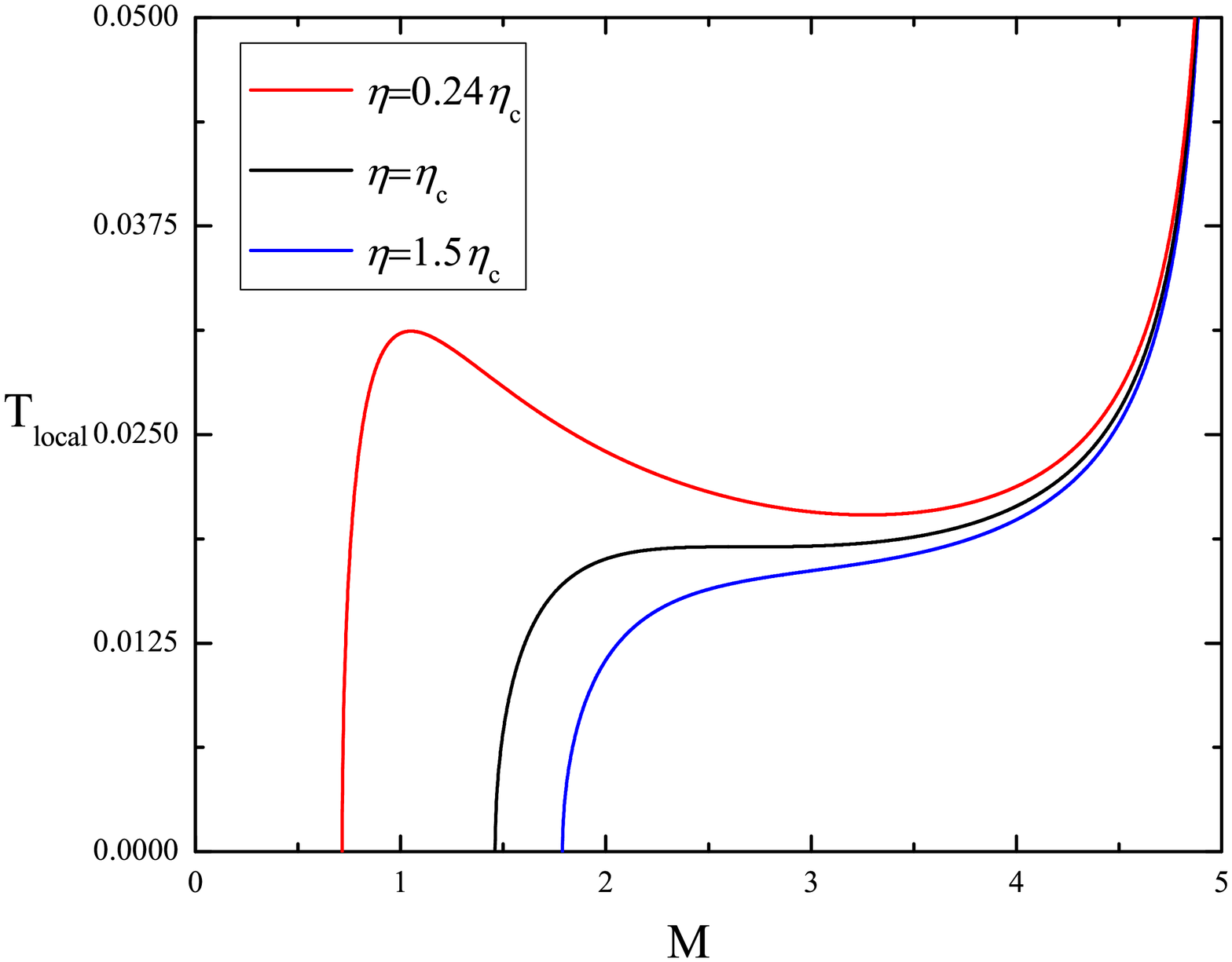}
\label{fig1-a}
\end{minipage}
}
\subfigure[]{
\begin{minipage}[b]{0.42\textwidth}
\includegraphics[width=1.25\textwidth]{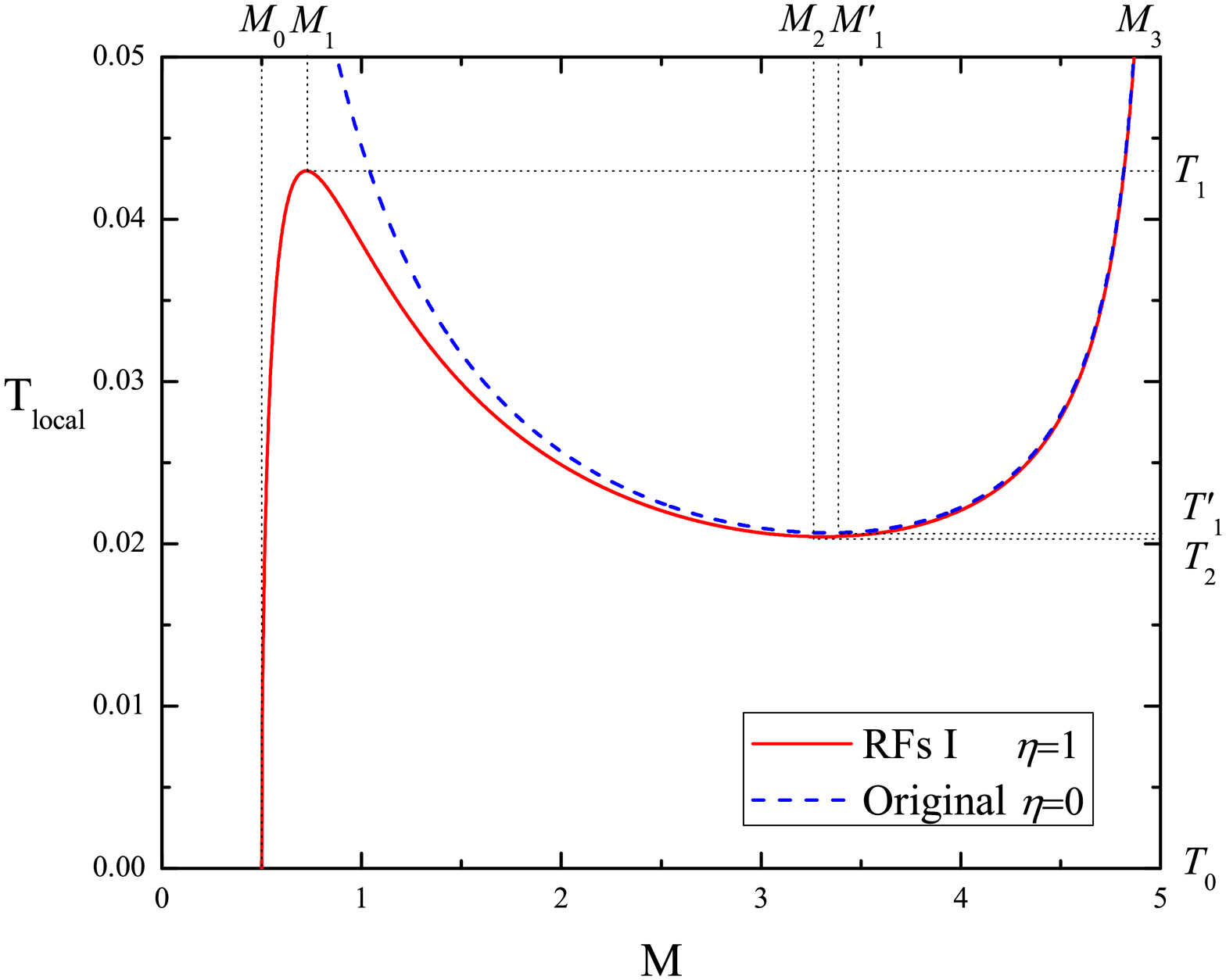}
\label{fig1-b}
\end{minipage}
}
\caption{(a) shows the relation between the rainbow local temperature and mass for different $\eta$.
(b) shows the original and rainbow local temperatures heat capacity versus mass. Here we set $ G = 1$ and  $r = 10$.}
\label{fig1}
\end{figure}

Now, let us calculate the local temperature of the rainbow SC black hole. Based on the identification in Refs.~\cite{ch42,ch66,ch66+,ch67+}, the local temperature in gravity's rainbow at a finite distance  $r$ outside the black hole can be expressed as
\begin{equation}
\label{eq9+}
T_{local}^{RFs} = T_H \left( {1 - \frac{{2GM}}{r}} \right)^{ - \frac{1}{2}} \frac{{g\left( {{E \mathord{\left/
 {\vphantom {E {E_p }}} \right.
 \kern-\nulldelimiterspace} {E_p }}} \right)}}{{f\left( {{E \mathord{\left/
 {\vphantom {E {E_p }}} \right.
 \kern-\nulldelimiterspace} {E_p }}} \right)}},
 \end{equation}
Considering Eq.~(\ref{eq8}) and the specific value for rainbow functions (\ref{eq2}), and putting $n=2$, the corresponding local temperature for the observer on the cavity is given by
\begin{equation}
\label{eq10}
T_{local}^{RFsI}  = \frac{1}{{16\pi \left( {GM} \right)^2 \sqrt {1 - \frac{{2GM}}{r}} }}\sqrt {\left( {2GM} \right)^2  - \eta G} .
\end{equation}
Obviously, the Eq.~(\ref{eq10}) is implemented by the redshift factor of the metric. If we take the parameter $r$ as a invariable quantity, the critical value of black hole's mass, the rainbow parameter and local temperature can be obtained by the following equations
\begin{equation}
\label{eq10+}
\left( {\frac{{\partial T_{local} }}{{\partial M}}} \right)_r  = 0, \quad   \left( {\frac{{\partial ^2 T_{local} }}{{\partial M^2 }}} \right)_r  = 0.
\end{equation}
Inserting Eq.~(\ref{eq10}) and setting $r=10$ as an example, it is found that the critical mass is $M_{c}^{RFsI}={8 \mathord{\left/ {\vphantom {8 3}} \right. \kern-\nulldelimiterspace} 3}$, the critical value of rainbow parameter is $\eta_{c}={{128} \mathord{\left/ {\vphantom {{128} {15}}} \right. \kern-\nulldelimiterspace} {15}}$ and the critical local temperature is $T_{c}^{RFsI} = {{3\sqrt {{3 \mathord{\left/ {\vphantom {3 2}} \right. \kern-\nulldelimiterspace} 2}} } \mathord{\left/ {\vphantom {{3\sqrt {{3 \mathord{\left/ {\vphantom {3 2}} \right. \kern-\nulldelimiterspace} 2}} } {64\pi }}} \right. \kern-\nulldelimiterspace} {64\pi }}$, respectively. Numerical computations drive us to the corresponding ``$T_{local}-M$'' diagram which is plotted in Fig.~\ref{fig1}. In Fig.~\ref{fig1-a}, it follows that for $\eta \neq 0$, there is a phase transition with $0 < \eta < \eta_{c}$. Therefore, for the convenience of discussing the critical phenomena and phase transition of rainbow SC black hole, we will set $\eta=1$ in the next calculation.

By fixing $G = 1$ and $r=10$, we plot the rainbow local temperature $T_{local}^{RFsI}$ and original case $T_{local}$ versus the mass of the black hole in Fig.~\ref{fig1-b}. The blue dashed line illustrates the original local temperature, which is infinite when the mass approaches zero and $M_3$. The minimum value of local temperature $T_1'$  occurs at  $M_1'$. For the rainbow local temperature case (red solid line), one can see that $T_{local}^{RFsI}$ recovers the original local temperature at  $M_3$, where the size of black hole approaches the horizon of SC black hole $r = 2GM_3$, it indicates that the black hole is very hot near the event horizon. As the mass of the black hole decreases, the rainbow local temperature reduces to $T_2$ corresponding to the mass $M_2$, and then increasing to its maximum value $T_1$ at $M_1$. At last, the rainbow local temperature goes to zero when the mass of black hole approaches a finite value, which leads to a remnant (namely, $M_{res} = M_0 = \sqrt {{\eta  \mathord{\left/ {\vphantom {\eta  {4G}}} \right. \kern-\nulldelimiterspace} {4G}}}$). Those results are consistent with Ali's analysis in Ref.~\cite{ch31}. The values of $\left( {M_1 ,T_1 } \right)$, $\left( {M_2 ,T_2 } \right)$, $\left( {M_3 ,T_3 } \right)$ and $\left( {M_1' ,T_1' } \right)$ can be numerically obtained if needed.

\begin{figure}[H]
\centering 
\includegraphics[width=.6\textwidth,origin=c,angle=0]{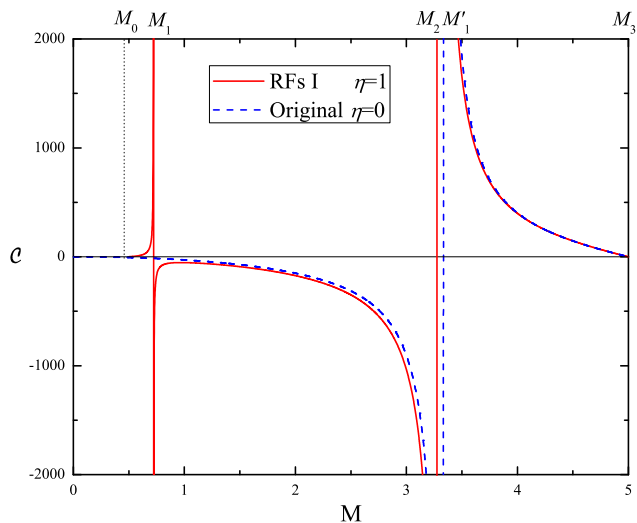}
\caption{\label{fig2} The original and rainbow heat capacity versus mass for $G = 1$ and  $r = 10$.}
\end{figure}

Next, using the thermodynamic first law, the total thermodynamic internal energy of rainbow SC black hole within the boundary $r$ is given by
\begin{equation}
\label{eq11}
E_{local}^{RFsI}  = \int_{M_0 }^M {T_{local}^{RFsI} dS^{RFsI} }  = \frac{r}{G}\left( {\sqrt {1 - \frac{{\sqrt {\eta G} }}{r}}  - \sqrt {1 - \frac{{2GM}}{r}} } \right),
\end{equation}
which goes to the local energy of original SC black hole $E_{local}  = {{r\left( {1 - \sqrt {1 - {{2GM} \mathord{\left/ {\vphantom {{2GM} r}} \right.
 \kern-\nulldelimiterspace} r}} } \right)} \mathord{\left/ {\vphantom {{r\left( {1 - \sqrt {1 - {{2GM} \mathord{\left/ {\vphantom {{2GM} r}} \right.
 \kern-\nulldelimiterspace} r}} } \right)} G}} \right. \kern-\nulldelimiterspace} G}$ when $\eta \rightarrow 0$. It may be noted that one can investigate  thermodynamic stability of the black holes through the heat capacity. Hence, employing the rainbow local temperature Eq.~(\ref{eq10}) and the local energy Eq.~(\ref{eq11}), the rainbow heat capacity at fixed $r$ can be expressed as
\begin{equation}
\label{eq12}
\mathcal{C}^{RFsI} =  {\frac{{\partial E_{local}^{RFsI} }}{{\partial T_{local}^{RFsI} }}}  = \frac{{16GM^3 \left( {2GM - r} \right)\chi }}{{4GM^2 \left( {r - 3GM} \right) + \eta (5GM - 2r)}}.
\end{equation}
By setting $\mathcal{C}^{RFsI} =0$, the remnant mass is $M_{res} = \sqrt {{\eta  \mathord{\left/ {\vphantom {\eta  {4G}}} \right. \kern-\nulldelimiterspace} {4G}}}$. The above equation reduces to the heat capacity of original SC black hole $ \mathcal{C}  = 8\pi GM^2 (r - 2GM)/(3GM - r)$ when  $\eta$ vanishes. The behaviors of the heat capacity can be observed from Fig.~\ref{fig2}.

In Fig.~\ref{fig2}, the red solid line for rainbow heat capacity $\mathcal{C}^{RFsI}$ vanishes at $M_0$, which is in the contrary to the blue dashed line for the original heat capacity $\mathcal{C}$ that goes to zero when $M \to 0$. It is well known that the black hole is stable if the heat capacity is positive, whereas the black hole is unstable if it with negative value. So that, we can find that the rainbow heat capacity has two stable regions of $M_0 \leq M \leq M_1$ and  $M_2 \leq M \leq M_3$, and one unstable regions of $M_1 \leq M \leq M_2$, while blue dashed line for the original case only has one stable regions of  $M_1' \leq M \leq M_3$ and one unstable regions of $ M \leq M_1'$. Interestingly, the rainbow heat capacity diverges at the points where the rainbow SC black hole temperature reaches its maximum value $M_1$ and minimum value $M_2$, which indicates there exists two second order phase transitions in the canonical ensemble.

Based on the discussions about the rainbow local temperature and the heat capacity, we classify the rainbow SC black hole to three branches according to its mass scale. The ranges, states and stability for three branches of the rainbow SC black hole are shown in Table.~\ref{tab1}.

\begin{table}[H]
\centering
\caption {\label{tab1} The ranges, states and stability for three branches of the SC black hole in the framework of RFs I.}
\begin{tabular}{c c c c c c}
\toprule
Branche   &         Range         &  State         & Stability \\
\midrule
1         &$ M_0 \leq M \leq M_1$ &   small        &  stable   \\
2         &$ M_1 \leq M \leq M_2$ &  intermediate  &  unstable  \\
3         &$ M_2 \leq M \leq M_3$ &   large        &  stable    \\
\bottomrule
\end{tabular}
\end{table}

From Table.~\ref{tab1}, one can find an additional intermediate black hole in the system, which never appears in the Hawking-Page phase transition. Besides, it is easy to see that the small black hole (SBH) and the large black hole (LBH) are stable, whereas the intermediate black hole (IBH) is unstable.

\begin{figure}[H]
\centering
\subfigure[]{
\begin{minipage}[b]{0.31\textwidth}
\includegraphics[width=1.2\textwidth]{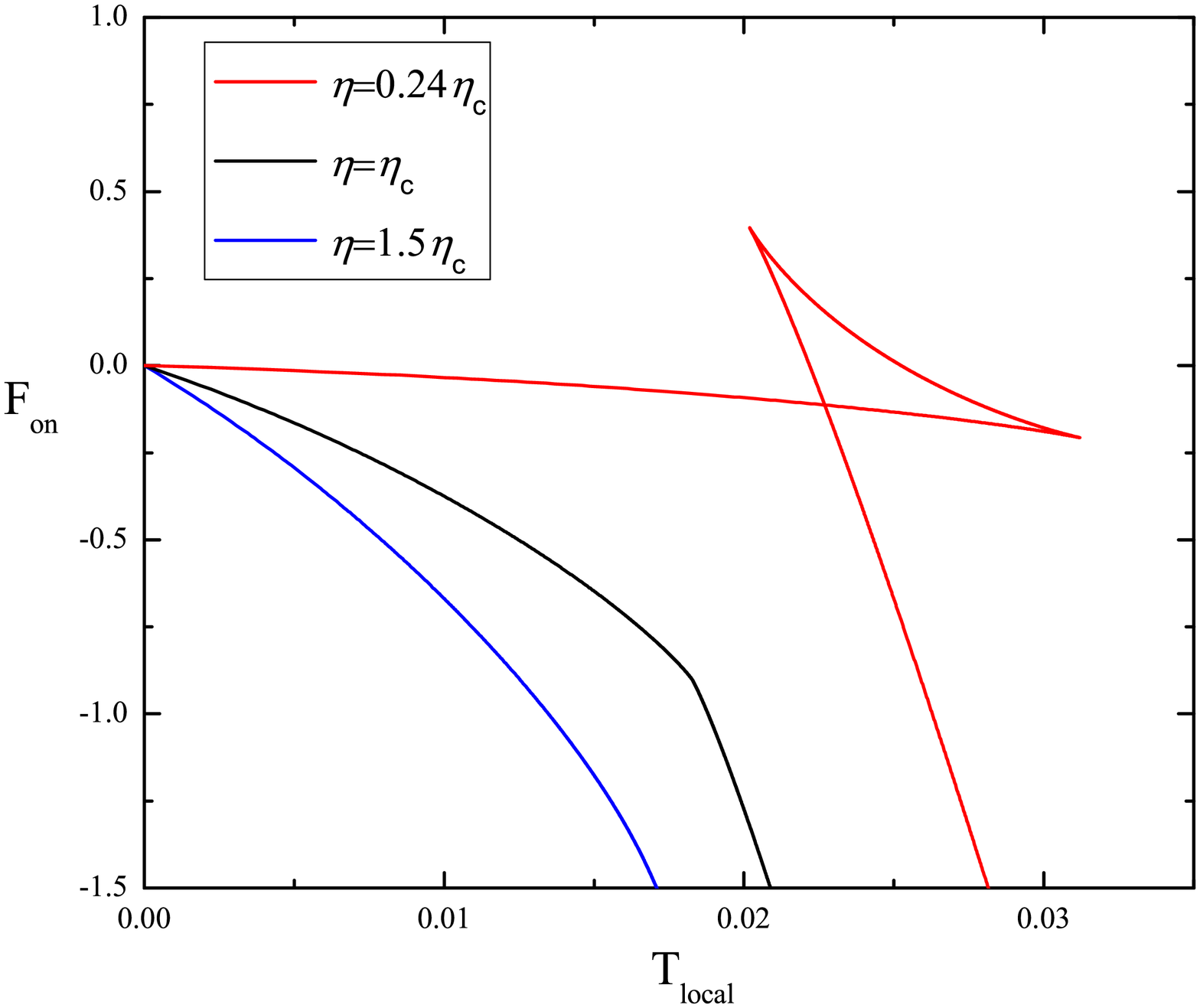}
\label{fig3-a}
\end{minipage}
}
\subfigure[]{
\begin{minipage}[b]{0.31\textwidth}
\includegraphics[width=1.2\textwidth]{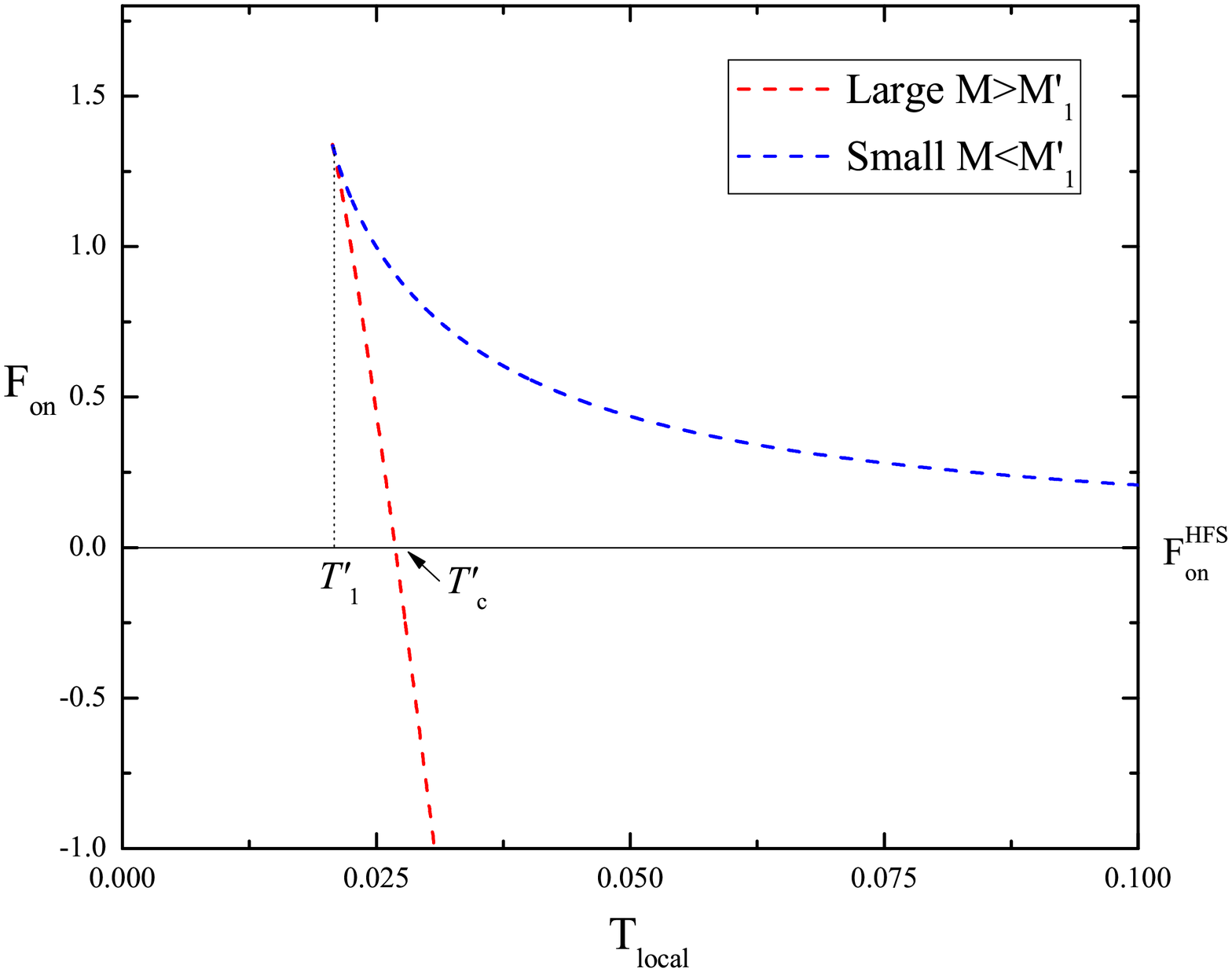}
\label{fig3-b}
\end{minipage}
}
\subfigure[]{
\begin{minipage}[b]{0.31\textwidth}
\includegraphics[width=1.2\textwidth]{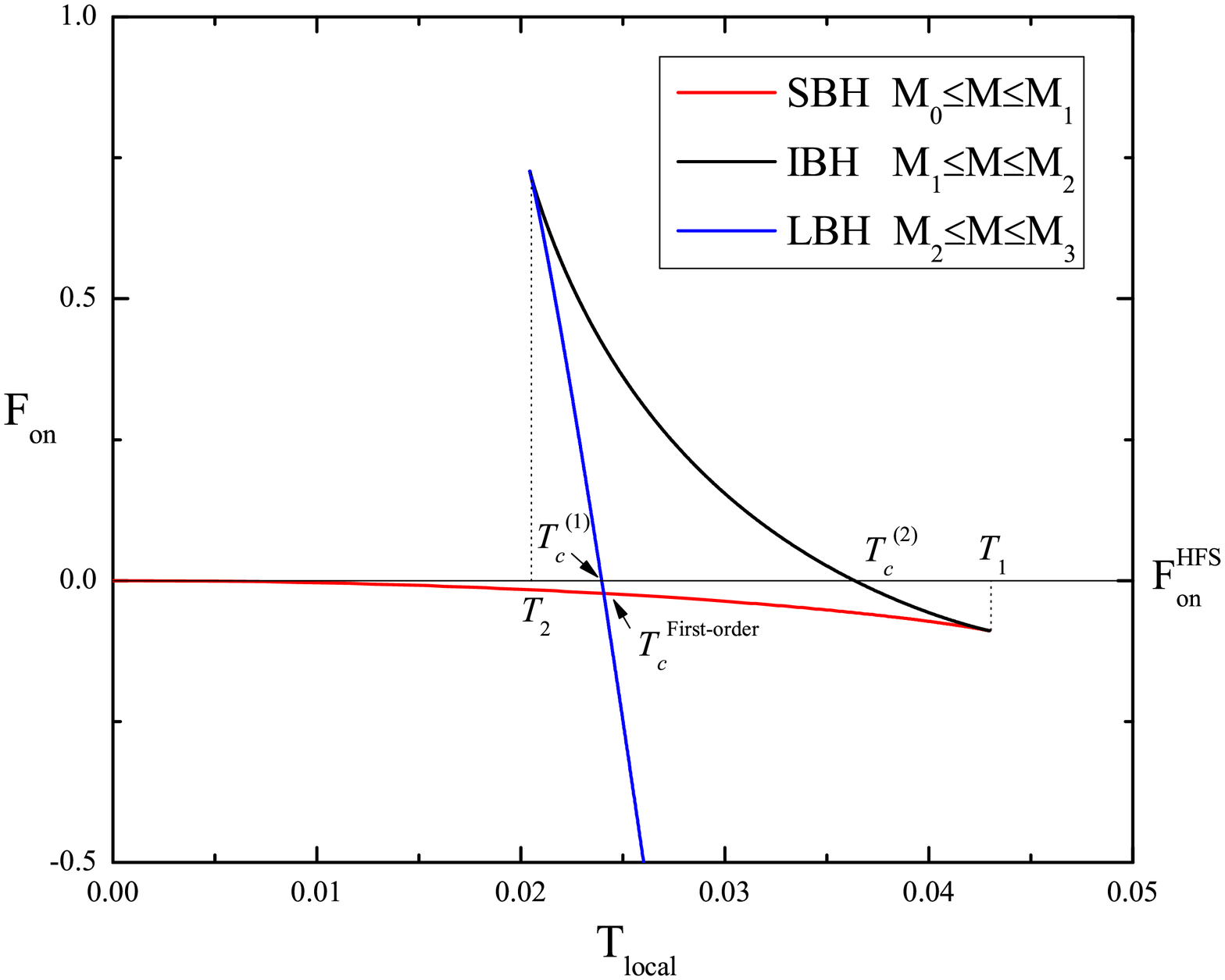}
\label{fig3-c}
\end{minipage}
}
\caption{For a rainbow SC black hole (a) shows the variation of free energy with the local temperature for different $\eta$.
(b) and (c) show the original and rainbow free energy of SC black hole as function of the local temperatures. Here we choose $G = 1$ and $r=10$.}
\label{fig3}       
\end{figure}

In order to obtain more details of the thermodynamic phase transition of the rainbow SC black hole, it is necessary to investigate the free energy of SC black hole in the gravity's rainbow enclosed in a cavity. In Refs.~\cite{ch67,ch68}, the free energy is defined as
\begin{equation}
\label{eq12+}
F_{on}  = E_{local}  - T_{local} S.
\end{equation}
Putting Eq.~(\ref{eq9}), Eq.~(\ref{eq10}) and Eq.~(\ref{eq11}) into Eq.~(\ref{eq12+}), the rainbow free energy is given by
\begin{equation}
\label{eq13}
F_{on}^{RFsI}  = \frac{r}{G}\left( {\sqrt {1 - \frac{{\sqrt {\eta G} }}{r}}  - \sqrt {1 - \frac{{2GM}}{r}} } \right) - \frac{\chi }{{\sqrt {1 - {{2GM} \mathord{\left/ {\vphantom {{2GM} r}} \right. \kern-\nulldelimiterspace} r}} }}\left[ {\frac{{2MG\chi  + \eta {\rm{ln}}\left( {2GM + \chi } \right)}}{{16G^2 M^2 }}} \right].
\end{equation}
Note that for $\eta=0$, the above equation is reduced to the original free energy  $ F_{on}  = r{{\left( {1 - \sqrt {1 - {{2GM} \mathord{\left/ {\vphantom {{2GM} r}} \right. \kern-\nulldelimiterspace} r}} } \right)} \mathord{\left/ {\vphantom {{\left( {1 - \sqrt {1 - {{2GM} \mathord{\left/ {\vphantom {{2GM} r}} \right. \kern-\nulldelimiterspace} r}} } \right)} G}} \right. \kern-\nulldelimiterspace} G} - {M \mathord{\left/ {\vphantom {M {2\sqrt {1 - {{2GM} \mathord{\left/ {\vphantom {{2GM} r}} \right. \kern-\nulldelimiterspace} r}} }}} \right. \kern-\nulldelimiterspace} {2\sqrt {1 - {{2GM} \mathord{\left/ {\vphantom {{2GM} r}} \right. \kern-\nulldelimiterspace} r}} }}$. For further investigate the phase transition between the black holes and the hot flat space (HFS), we plot Fig.~\ref{fig3}.

Fig.~\ref{fig3-a} shows $F_{on}$ curves with $T_{local}$ for different $\eta$, the value of rainbow parameter $\eta$ decreases from top to bottom. It is clear that the swallow tail structure appears when the rainbow parameter $\eta$ is smaller than the critical value $\eta_{c}^{RFsI}$, which indicates there is a two-phase coexistence state. The results are consistent with the profile of $T_{local}-M$ in Fig.~\ref{fig1-a}.

As seen from the middle and right panels of Fig.~\ref{fig3}, the Fig.~\ref{fig3-b} shows the free energy of original SC black hole and the Fig.~\ref{fig3-c} depicts the free energy of  rainbow SC black hole, the $F_{on}^{{\rm{HFS}}}$  represents the free energy of the hot flat space. First, let us focus on the original case. It is found that the free energy vanishes when $M \to {\rm{0}}$  since the vacuum state is Minkowski space-time.  The small-large (Hawking-Page) transition occurs at $T_1'$. For  $T < T_c ^\prime$, both the small and large black holes are higher than the $F_{on}^{{\rm{HFS}}}$, it means that the HFS is more probable than the small and large black holes. However, for  $T > T_c ^\prime$, the small black hole higher is than the $F_{on}^{{\rm{HFS}}}$ while the large black hole is lower than the $F_{on}^{{\rm{HFS}}}$, which indicates that the large black hole is more probable than the HFS. Therefore, one can find a phase transition in this thermodynamic system above $T_c ^\prime$, which leads to the radiation collapse to the large black hole and the small black hole eventually decays into the large black hole thermodynamically.

Now, turn to look at Fig.~\ref{fig3-c}. Evidently, the free energy of the hot flat space is $F_{on}^{{\rm{HFS}}}=0$ which is coincide with the horizontal coordinate. For non-zero  $\eta$, the behavior of the rainbow free energy is apart from the original picture.

(i) The small-intermediate black hole transition occurs at the inflection point $T_1$  corresponding to the mass  $M_1$, the intermediate and large black holes are degenerate at $T_2$  corresponding to the mass $M_2$. Besides, it is easy to find that the free energy of large black hole (blue solid line) and the $F_{on}^{{\rm{HFS}}}$ intersect at Hawking-Page-type critical temperature $T_c^{(1)}$, the free energy of intermediate black hole (black solid line) and the $F_{on}^{{\rm{HFS}}}$ intersect at another Hawking-Page-type critical temperature $T_c^{(2)}$.

(ii) There exists a first order phase transition since the $F_{on}$ demonstrates the characteristic swallow tail behavior. It is reminiscent of the ``free energy-Hawking temperature'' relation of charged AdS black hole in Refs.~\cite{ch51,ch52,ch53,ch54,ch55,ch56,ch57}. The intersection point between the red solid line and the blue solid line is the first order phase transition point that corresponding to the temperature $T_c^{First-order}$.

(iii) For  $T_0  < T < T_c^{(1)}$, the intermediate black hole and large black hole are higher than the  $F_{on}^{{\rm{HFS}}}$ while the small black hole is lower than $F_{on}^{{\rm{HFS}}}$, hence the stable small black hole is more probable than the HFS. For  $T_c^{(1)}  < T < T_c^{First-order}$, both small and large black holes are lower than the the $F_{on}^{{\rm{HFS}}}$. However, the large black hole should decay into the small black hole since its free energy is higher than the free energy of the small black hole in this region. Consequently, at $T_0 < T < T_c^{First-order}$ the small black hole should undergoes a tunneling and the HFS can collapse to the small black hole. As the local temperature increase, one can find that free energies of black holes satisfy the relation $F_{on}^{{\rm{LBH}}} < F_{on}^{{\rm{SBH}}} < F_{on}^{{\rm{HFS}}} < F_{on}^{{\rm{IBH}}}$ for  $T_c^{First-order} < T < T_c^{\left( 2 \right)}$, whereas the relation becomes $F_{on}^{{\rm{LBH}}} < F_{on}^{{\rm{SBH}}} < F_{on}^{{\rm{IBH}}} < F_{on}^{{\rm{HFS}}} $ for  $T_c^{\left( 2 \right)}  < T < T_1$, it means that hot flat space does not only collapse into the small black hole but also into the intermediate and large black holes. However, the intermediate black hole with negative heat capacity is unstable, so that it would decay into the small black hole. Meanwhile, it is obvious that the free energy of small black hole is always higher than free energy of the large black hole, it leads to the small black hole eventually decays into the large black hole thermodynamically. Therefore, for  $T_c^{First-order}  < T < T_1$, the HFS would eventually decays into the large black hole due to the tunneling effect.

\section{Thermodynamics and phase transition of rainbow SC black hole with rainbow functions II}
\label{TP II}
\begin{figure}[H]
\centering
\subfigure[]{
\begin{minipage}[b]{0.42\textwidth}
\includegraphics[width=1.25\textwidth]{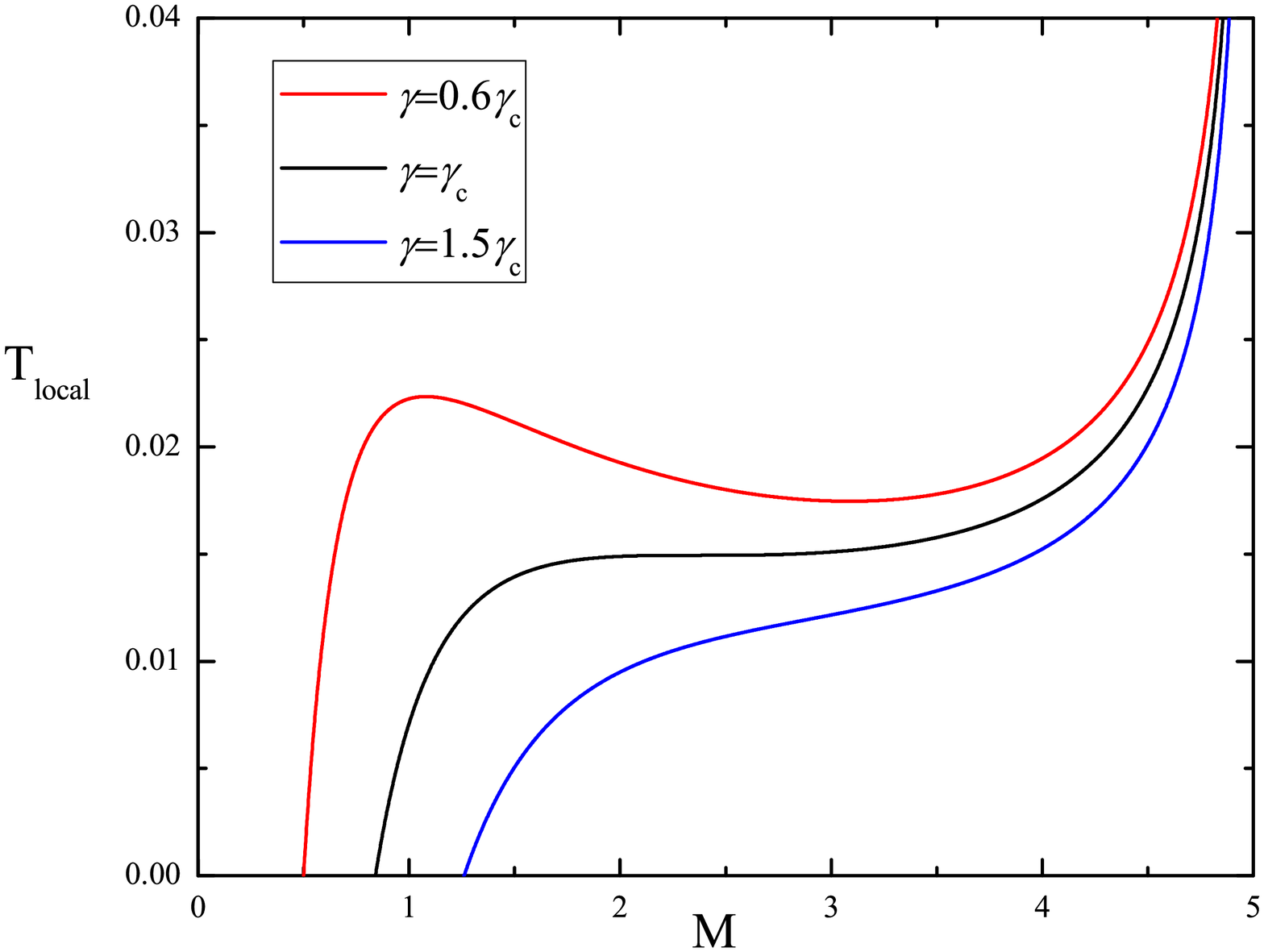}
\label{fig5-a}
\end{minipage}
}
\subfigure[]{
\begin{minipage}[b]{0.42\textwidth}
\includegraphics[width=1.25\textwidth]{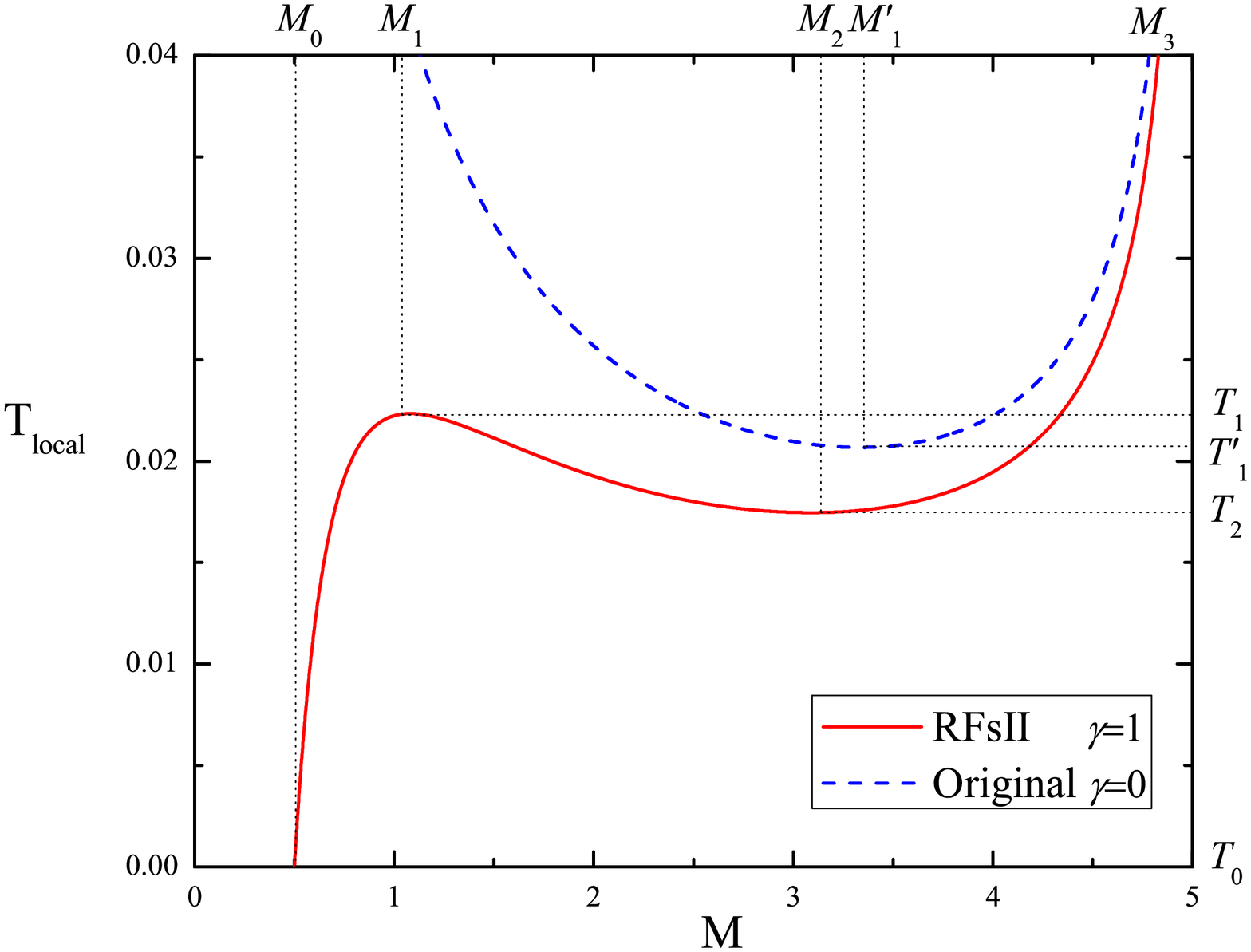}
\label{fig5-b}
\end{minipage}
}
\caption{(a) shows the relation between the rainbow local temperature and mass for different $\gamma$.
(b) shows The original and modified local temperature versus mass. We set $ G = 1$ and  $r = 10$.}
\label{fig5}      
\end{figure}

In this section, using Eq.~(\ref{eq3}), the thermodynamics and phase transition of SC black hole in the framework of RFs II will be investigated. A substitution in Eq.~(\ref{eq5}) gives the following result
\begin{equation}
\label{eq13}
T_H^{RFsII}  = \frac{1}{{8\pi GM}}\left( {1 - \frac{{\gamma  }}{{2 \sqrt G M}}} \right).
\end{equation}
It should be noted that the relation of lower bound of the energy $E \ge {1 \mathord{\left/ {\vphantom {1 {r_H }}} \right. \kern-\nulldelimiterspace} {r_H }} = {1 \mathord{\left/ {\vphantom {1 {2GM}}} \right. \kern-\nulldelimiterspace} {2GM}}$  and the natural unit  $E_p  = {1 \mathord{\left/{\vphantom {1 {\sqrt G }}} \right. \kern-\nulldelimiterspace} {\sqrt G }}$ are used again for obtaining the above expression. With the help of the first law of thermodynamics, the new rainbow entropy is given by
\begin{equation}
\label{eq14}
S^{RFsII}  = \int {T_H^{ - 1} dM = 4\sqrt G M\pi \left( {\sqrt G M + \gamma } \right) + 2\pi \eta ^2 {\rm{ln}}\left( {2\sqrt G M - \gamma } \right)} ,
\end{equation}
which becomes the original entropy $S = {A \mathord{\left/ {\vphantom {A 4}} \right. \kern-\nulldelimiterspace} 4}$ when $\gamma$ vanishes. Next, according to Eq.~(\ref{eq9+}), one can obtain the local temperature
\begin{equation}
\label{eq15}
T_{local}^{RFsII}  = \frac{1}{{8\pi GM\sqrt {1 - \frac{{2GM}}{r}} }}\left( {1 - \frac{\gamma }{{2\sqrt G M}}} \right).
\end{equation}
The above equation leads to the existence of a minimum mass below which the local temperature becomes a negative quantity $M_0 = {{\gamma \sqrt G } \mathord{\left/ {\vphantom {{\eta \sqrt G } 2}} \right. \kern-\nulldelimiterspace} 2}$. Using Eq.~(\ref{eq10+}) and setting $r=10$, the critical mass, critical free parameter and critical local temperature can be expressed as follows
\begin{equation}
\label{eq15+}
M_{c}^{RFsII}=2.36701, \qquad \gamma_{c}=1.68082, \qquad T_{c}^{RFsII}=0.0149398.
\end{equation}
It shows that the phase transition appears when $0 < \gamma < \gamma_{c}$. This is depicted in Fig.~\ref{fig5-a}. In order for the study to be focused and feasible, we take the free parameter as $\gamma = 1$ in the next discussion.

Now, let us plot the original and rainbow local temperature versus mass for $ G = 1$ and  $r = 10$ in Fig.~\ref{fig5-b}. It shows that the blue dashed line represents the local temperature of original SC black hole, and the red solid line represents the local temperature of rainbow black hole. The evaporation process of SC black hole in the framework of RFs II is commonly organized in three stages: at the first stage ($M_2  \leq M \leq M_3$), the rainbow local temperature decreases through its evaporation process. At the second stage ($M_1 \leq M \leq M_2$), the rainbow local temperature increases as mass decreases, lasting up to the mass in which it comes near to a maximum value. At the final stage ($M_0 \leq M \leq M_1$), the mass of rainbow SC black hole cannot get smaller than $M_0$ since the negative temperature becomes violates the laws of thermodynamics, that is, the remnant mass is $M_0=M_{res}={{\gamma \sqrt G } \mathord{\left/ {\vphantom {{\eta \sqrt G } 2}} \right. \kern-\nulldelimiterspace} 2}$. Hence, it has no physical meaning. Obviously, the behavior of the rainbow local temperature of Eq.~(\ref{eq15}) is similar as that of Eq.~(\ref{eq10}) as long as  $M_0 \leq M \leq M_3$. It means that the RFs II can stop the Hawing radiation and leads to a remnant of black hole. Therefore, we can also define large black hole (LBH) where $M_1  \leq M \leq M_2$, intermediate black hole (IBH) where $M_2  \leq M \leq M_3$, and small black hole (SBH) $M_0 \leq M \leq M_1$. Notedly, there is only one small black hole for $T_0 \leq T \leq T_2$, one large black hole for $ T \geq T_1$, and all the three states for $T_2  \leq T \leq T_1$. The similar results can be found in Refs.~\cite{ch69,ch70,ch71,ch72,ch73,ch74,ch75}.

\begin{figure}[H]
\centering 
\includegraphics[width=.6\textwidth,origin=c,angle=0]{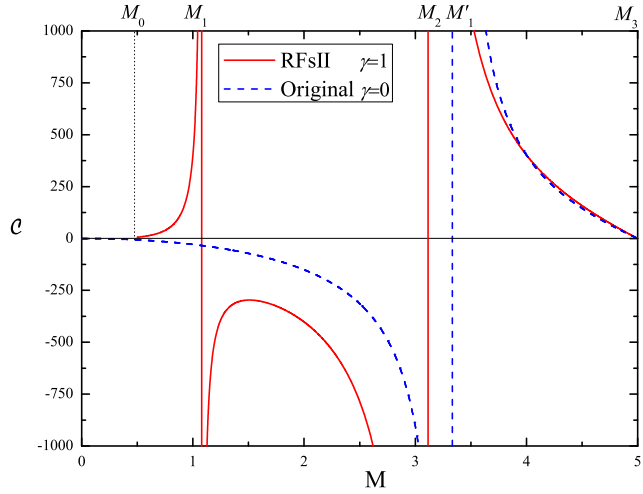}
\caption{\label{fig6} The original and modified heat capacity versus mass for $ G = 1$ and  $r = 10$.}
\end{figure}

With the help of the expression of total internal energy Eq.~(\ref{eq15}), one can easily obtain the result $E_{local}^{RFsII}  = \int_{M_0 }^M {T_{local}^{RFsII} dS^{RFsII} }  = {{r\left( {\sqrt {1 - {{\gamma G^{{3 \mathord{\left/ {\vphantom {3 2}} \right. \kern-\nulldelimiterspace} 2}} } \mathord{\left/ {\vphantom {{\gamma G^{{3 \mathord{\left/ {\vphantom {3 2}} \right. \kern-\nulldelimiterspace} 2}} } r}} \right. \kern-\nulldelimiterspace} r}}  - \sqrt {1 - {{2GM} \mathord{\left/ {\vphantom {{2GM} r}} \right. \kern-\nulldelimiterspace} r}} } \right)} \mathord{\left/ {\vphantom {{r\left( {\sqrt {1 - {{\gamma G^{{3 \mathord{\left/ {\vphantom {3 2}} \right. \kern-\nulldelimiterspace} 2}} } \mathord{\left/ {\vphantom {{\gamma G^{{3 \mathord{\left/
 {\vphantom {3 2}} \right. \kern-\nulldelimiterspace} 2}} } r}} \right. \kern-\nulldelimiterspace} r}}  - \sqrt {1 - {{2GM} \mathord{\left/
 {\vphantom {{2GM} r}} \right. \kern-\nulldelimiterspace} r}} } \right)} G}} \right. \kern-\nulldelimiterspace} G}$ which is the same as Eq.~(\ref{eq11}). Then, using again expression Eq.~(\ref{eq12}), the rainbow heat capacity at fixed $r$ is given as
\begin{eqnarray}
\label{eq16}
\mathcal{C}^{RFsII}  = \left( {\frac{{\partial E_{local}^{RFsII} }}{{\partial T_{local}^{RFsII} }}} \right)_r = \frac{{16M^3 \pi \left( {r - 2GM} \right)}}{{2M\left( {3GM - r} \right) + \gamma \sqrt G \left( {2r - 5GM} \right)}}.
\end{eqnarray}
The variation of the heat capacity  $\mathcal{C}^{RFsII}$ with the mass $M$  is plotted in Fig.~\ref{fig6}.

As seen in Fig.~\ref{fig6}, the heat capacity of SC black hole obeying a MRD that based on the RFs II (red solid line) changes its sign at $M_1$  and $M_2$  for which the denominator in Eq.~(\ref{eq16}) vanishes. Meanwhile, the rainbow heat capacity $\mathcal{C}^{RFsII}$  goes to zero when mass approaches to $M_0$, which is similar as what we found in the RFs I case. When $M \to M_3$, the rainbow heat capacity tends to the original case (blue dashed line) since the effect of quantum gravity is negligible at that point. The rainbow heat capacity is divergent at the inflection points $M_1$  and $M_2$. Hence, the phase transitions near there are second orders. Adopting the similar strategy, the states of SC black hole can be classifies to three branches according to the mass scale, which are shown in Table~\ref{tab2}.

\begin{figure}[H]
\centering
\subfigure[]{
\begin{minipage}[b]{0.42\textwidth}
\includegraphics[width=1.25\textwidth]{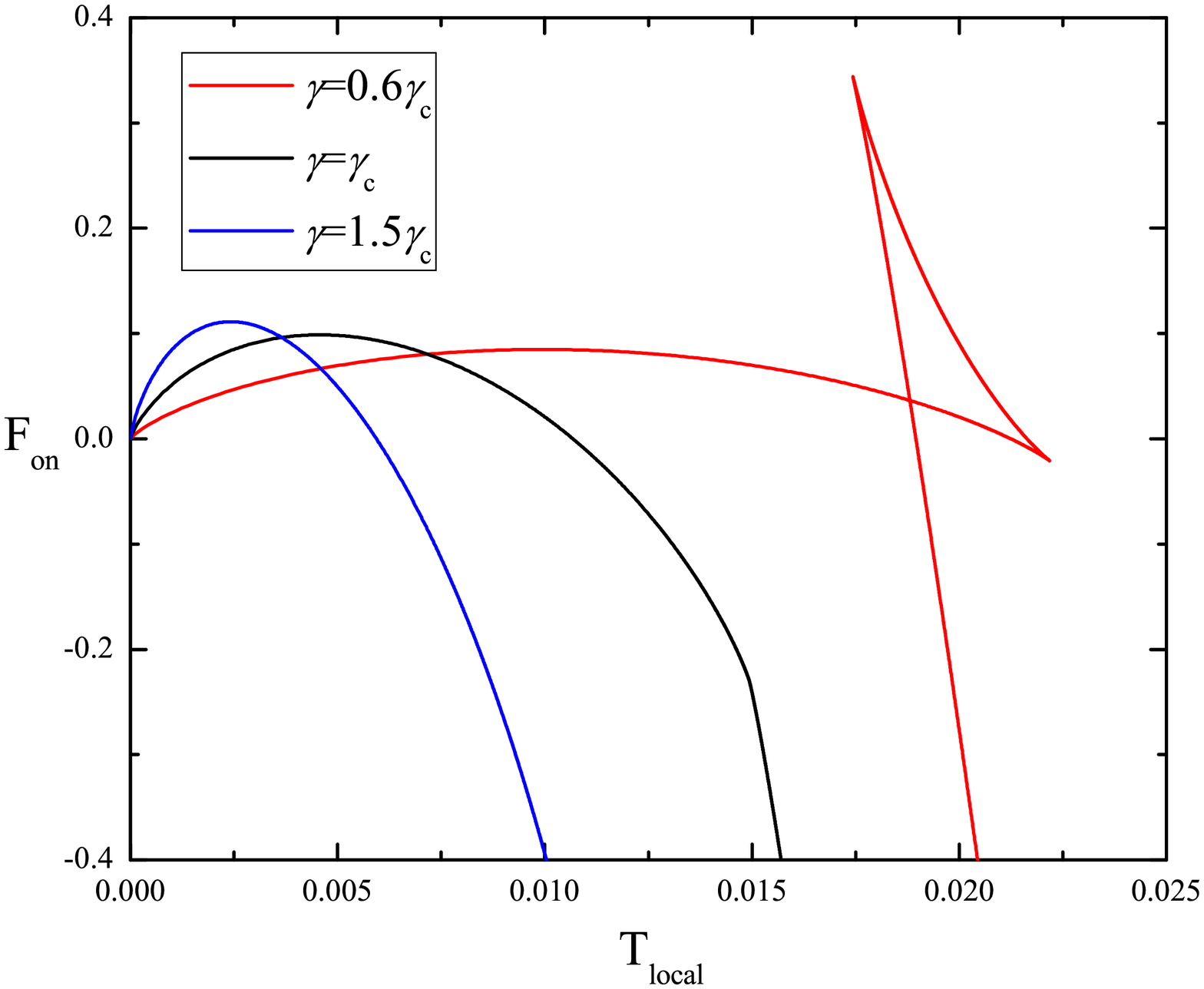}
\label{fig7-a}
\end{minipage}
}
\subfigure[]{
\begin{minipage}[b]{0.42\textwidth}
\includegraphics[width=1.25\textwidth]{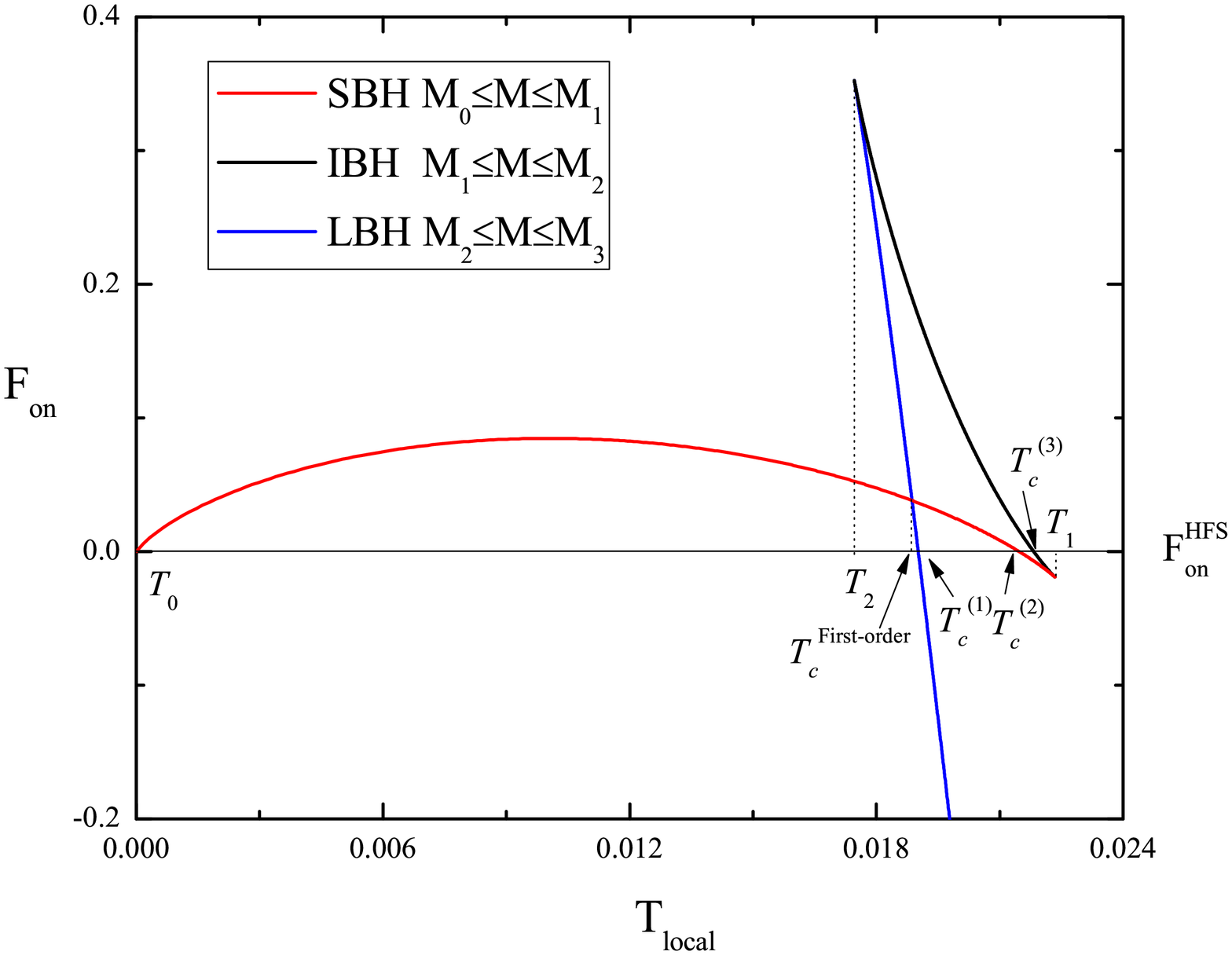}
\label{fig7-b}
\end{minipage}
}
\caption{For a rainbow SC black hole (a) shows the variation of free energy with the local temperature for different $\eta$.
(b) shows the original and modified free energy of SC black hole as function of the local temperature. Here we choose $G = 1$ and $r=10$.}
\label{fig7}      
\end{figure}

\begin{table}[htbp]
\centering
\caption {\label{tab2} The ranges, states and stability for three branches of the SC black hole in the framework of RFs II.}
\begin{tabular}{c c c c c c}
\toprule
Branche   &         Range         &  State       & Stability \\
\midrule
1         &$M_0 \leq M \leq M_1 $ &   small      &  stable   \\
2         &$ M_1 \leq M \leq M_2$ & intermediate &  unstable  \\
3         &$ M_2 \leq M \leq M_3$ &   large      &  stable    \\
\bottomrule
\end{tabular}
\end{table}
It is obvious that the results in Table~\ref{tab2} are similar as those in Table~\ref{tab1}. In the next discussion, we will use the three states of black hole to analyze the phase transition of rainbow SC black hole. Now, based on the relation  $F_{on}  = E_{local}  - T_{local} S$, the free energy is given by
\begin{eqnarray}
\label{eq17}
F_{on}^{RFsII}  &=  &\frac{r}{G}\left( {\sqrt {1 - \frac{{\gamma G^{{3 \mathord{\left/ {\vphantom {3 2}} \right. \kern-\nulldelimiterspace} 2}} }}{r}}  - \sqrt {1 - \frac{{2GM}}{r}} } \right)
\nonumber \\
& - & \frac{{\left( {2M - \sqrt G \gamma } \right)\left[ {2M\left( {M + \sqrt G \gamma } \right) + G\gamma ^2 {\rm{ln}}\left( {{{2M} \mathord{\left/ {\vphantom {{2M} {\sqrt G }}} \right. \kern-\nulldelimiterspace} {\sqrt G }} - \gamma } \right)} \right]}}{{8M^2 \sqrt {1 - \frac{{2GM}}{r}} }}.
\end{eqnarray}
Note that the above equation recovers the original free energy for $\gamma = 0$. Besides, as seen from Fig.~\ref{fig7-a} that the thermodynamic phase transition of the rainbow SC black hole occurs for $0 < \gamma< \gamma_{c}^{RFs II}$, which is consistent with the results in Fig.~\ref{fig5-a}. Next, by setting $ G = 1$ and  $r = 10$, one can see the behavior of the modified free energy $F_{on}^{RFsII}$ versus local temperature $T_{local}$ in Fig.~\ref{fig7-b}. One can find that

(i) The tiny-small black hole transition occurs at the inflection point $T_1$  corresponding to  $M_1$, and the small black hole and large black hole are degenerate at $T_2$  with  $M_2$. Interestingly, it can be easily found three intersections (Hawking-Page-type critical temperature $T_{c}^{(1)}$, $T_{c}^{(2)}$ and $T_{c}^{(3)}$) between the line of free energy and the $F_{on}^{{\rm{HFS}}}$ in Fig.~\ref{fig7-b} while there is only one intersection in Hawking-Page phase transition case as seen from Fig.~\ref{fig3-a} and two intersections in RFs I case as seen from Fig.~\ref{fig3-b}.

(ii) Evidently, there is a order phase transition at critical temperature $T^{First-order}_{c}$ since the images shows a characteristic swallow tail behavior.

(iii) As seen from Fig.~\ref{fig7-b}, all the three states of rainbow SC black hole are higher than the free energy of the hot flat space for $T_0 < T < T_{c}^{(1)}$, which means that the HFS is more probable in this region. However, the free energies of LBH, SBH and IBH drop below the $ F_{on}^{HFS}$ one by one above the $T_{c}^{(1)}$. In $T_{c}^{(1)} < T < T_{c}^{(2)}$, the relation of free energies obey $F_{on}^{{\rm{LBH}}} < F_{on}^{{\rm{HFS}}} < F_{on}^{{\rm{SBH}}} < F_{on}^{{\rm{IBH}}} $. Then, it changes to $F_{on}^{{\rm{LBH}}} < F_{on}^{{\rm{SBH}}}  < F_{on}^{{\rm{HFS}}} < F_{on}^{{\rm{IBH}}} $ for $T_{c}^{(2)} < T < T_{c}^{(3)}$. Finally, the relation becomes $F_{on}^{{\rm{LBH}}} < F_{on}^{{\rm{SBH}}} < F_{on}^{{\rm{IBH}}} < F_{on}^{{\rm{HFS}}}$ for $T_{c}^{(3)} < T < T_1$. So that, the hot flat space decays into the stable large black hole eventually for $T_{c}^{(1)} < T < T_1$. Those difference are caused by the different rainbow functions.

\section{Discussion}
\label{Dis}
In this paper, using two different rainbow functions, we studied the quantum corrections to the thermodynamics and the phase transitions of the SC black hole. First of all, according to the rainbow surface gravity and the uncertainty principle, we calculated the rainbow Hawking temperature. Subsequently, the others modified thermodynamic quantities such as entropy, heat capacity as well as the local temperature are obtained by some manipulations. Finally, based on those modifications, we derived the critical points of the black hole thermodynamic ensemble and analyzed the thermodynamic stability and phase transition of the rainbow SC black hole.

It is clear that our results are different from the those of Hawking-Page phase transition. Meanwhile, comparing the results of RFs I case with those of RFs II case, we found they have some similarities. Firstly, both the two kinds of rainbow functions can stop the Hawking radiation in the end final stages of black holes' evolution and leads to black hole remnants, which are agree with the predictions of generalized  uncertainty principle (GUP). Secondly, as seen from Fig.~\ref{fig1-a} and Fig.~\ref{fig5-a}, the phase transition occur when the rainbow parameters are smaller than the critical points. So that, we have set the rainbow parameters as $\eta=\gamma=1$ in order for the study to be focused and feasible. Thirdly, from Eq.~(\ref{eq12}) and Eq.~(\ref{eq16}), one can see that there are two second order phase transitions since the heat capacity enjoy two divergencies at $M_1$ and $M_2$. Fourthly, according to the ``$F_{on}-T_{local}$'' plane, it is found there is an unstable black hole interpolating between the small stable black hole and large stable black hole. Finally, the $F_{on}$ surface demonstrates the characteristic swallow tail behavior, which implies the system of rainbow SC black hole has a first order transition. This behavior is reminiscent of the ``free energy-Hawking temperature'' relation of AdS black holes. It implies that the rainbow spacetimes and the AdS spacetimes may related to some extent. Furthermore, our calculations showed some differences between the RFs I case and RFs II. For the RFs II case, one may found three Hawking-Page-type critical points while there are two Hawking-Page-type critical points for RFs I case. Besides, the hot flat space in $T_{c}^{(1)}< T < T_1$ only decays into the stable large black hole for RFs II case. However, for RFs I case, the hot flat space in $T_0 < T < T_1$ decays into stable small black hole or the stable large black hole.

In our previous work, we found that the thermodynamics of rainbow SC black hole is similar as the GUP corrected thermodynamics of SC black hole. Therefore, It will be worthwhile to verify if our conclusions in this paper would still hold for the GUP corrected thermodynamics of SC black hole. We left this issues for forthcoming work.

\vspace*{3.0ex}
{\bf Acknowledgements}
\vspace*{1.0ex}
This work is supported by the Natural Science Foundation of China (Grant No. 11573022).


\end{document}